\documentclass[conference, 10pt]{IEEEtran}
\usepackage{cite}
\IEEEoverridecommandlockouts
\usepackage{epsfig}
\usepackage{amsmath}
\usepackage{theorem}
\usepackage{amsmath}
\usepackage{color}
\usepackage{mathrsfs}
\usepackage{amsfonts}
\usepackage[algo2e]{algorithm2e}
\usepackage{graphicx}
\usepackage{romannum}
\usepackage{multirow}
\usepackage{cite}
\usepackage{amsmath,amssymb,amsfonts}
\usepackage{algorithm}
\usepackage[noend]{algpseudocode}
\usepackage{graphicx}
\usepackage{textcomp}
\usepackage{pdfpages}
\usepackage{amssymb}
\usepackage{mathrsfs}
\usepackage{euscript}
\usepackage{graphicx}
\usepackage{multirow}
\usepackage{cite}
\usepackage{amsmath,amssymb,amsfonts}
\usepackage{graphicx}
\usepackage{textcomp}
\usepackage{subcaption}

\def\BibTeX{{\rm B\kern-.05em{\sc i\kern-.025em b}\kern-.08em
    T\kern-.1667em\lower.7ex\hbox{E}\kern-.125emX}}
\begin{document}

\title{Performance Trade-off Between Uplink and Downlink in Full-Duplex Communications}
\author{\IEEEauthorblockN{Ata Khalili$^{\dag}$,~Mohammad Robat Mili$^{\ddag}$,~and Derrick Wing Kwan Ng$^{\star}$}
$^{\dag}$Department of Electrical and Computer Engineering,
Tarbiat Modares University, Tehran, Iran\\$^{\ddag}$Department of Telecommunications and information processing,~Ghent University,~Belgium\\
$^{\star}$School of Electrical Engineering and Telecommunications,~The University of New South Wales, Australia\\
Email:~ata.khalili@ieee.org,~mohammad.robatmili@ieee.org,~w.k.ng@unsw.edu.au
\thanks{D. W. K. Ng is supported by funding from the UNSW Digital Grid Futures Institute, UNSW, Sydney, under a cross-disciplinary fund scheme and by the Australian Research Council's Discovery Project (DP190101363).}}
\maketitle
\begin{abstract}
In this paper, we formulate two multi-objective optimization problems (MOOPs) in orthogonal frequency-division
multiple access (OFDMA)-based
in-band full-duplex (IBFD) wireless communications.~The aim of this study is to exploit the performance trade-off between uplink and downlink where a wireless radio
simultaneously transmits and receives in the same frequency.~We consider maximizing the system throughput as the first MOOP and minimizing the system aggregate power consumption as the second MOOP between uplink and downlink,~while taking into account the impact of self-interference~(SI)~and quality of service provisioning.~We study the throughput and the transmit power trade-off between uplink and downlink via solving these two problems.~Each MOOP is a non-convex mixed integer non-linear programming~(MINLP)~which is generally intractable.
In order to circumvent this difficulty, a penalty function is introduced to reformulate the problem into a mathematically
tractable form.~Subsequently,~each MOOP is transformed into a single-objective optimization problem~(SOOP)~via the weighted Tchebycheff method
which is addressed by majorization-minimization~(MM)~approach.
Simulation results demonstrate
an interesting trade-off between the considered competing objectives.
\end{abstract}

\section{Introduction}
The fifth-generation wireless communication is a major breakthrough for realizing pervasive wireless communication systems
to meet the demands for ultra-high data rate~\cite{1,Multiple_Antenna_5G,Massive_Access_5G}.~In particular,~fulfilling such requirement needs more dedicated spectral resources for mobile communication services.
On the other hand,~the paucity of spectrum resources has turned resource allocation to one of the most challenging problems
in wireless communication design \cite{Massive_Access_5G}.~To address this issue,~in-band full duplex~(IBFD)~is considered as a promising technology providing a viable solution to tackle this challenge.
In particular,~utilizing wireless transmitters and receivers simultaneously over the same frequency band in IBFD communication systems offers the potential to double the spectral efficiency than that of existing communication systems~\cite{2}.
In this regard,~resource allocation design for simultaneous downlink and uplink transmission plays a key role to exploit the
promising performance gain brought by IBFD communication.

Recently,~there are a plethora of works focusing on resource allocation
strategies in orthogonal frequency division multiple access (OFDMA) to harness the interference in full-duplex (FD) netwroks,~thereby increases the
overall performance of such networks~\cite{3,4,5,6,7,Ata_G,Ata_T}.
 In \cite{3},~assignment of a transmission mode,~resource block allocation,~and power adjustment were investigated where a sub-optimal iterative solution was proposed.
 In \cite{4}, a game theoretic approach was adopted to maximize the total system data rate.~However, the  quality   of   service~(QoS)~requirements of users were not taken into account.
Besides,~although the work in \cite{5} investigated the power control in IBFD communication in order to maximize the system throughput, the sub-channel assignment was not designed.
In \cite{6}, the resource allocation for multiuser networks incorporating a~(FD) base station (BS) and half-duplex~(HD)~users in non-orthogonal multiple access was studied.~The authors of \cite{Ata_G}~proposed a multi-objective resource allocation design for IBFD systems to strike a balance between energy efficiency and spectral efficiency.~The joint sub-channel assignment and power allocation for energy efficiency maximization in IBFD communication was considered in \cite{Ata_T} where an iterative algorithm based on Dinkelbach method was adopted to solve the considered design problem.~In \cite{7},~the authors proposed a framework to minimize the aggregate power consumption in uplink and downlink
simultaneously for secure multi-user wireless communication systems.~Nevertheless,~optimal subchannel assignment was not investigated and the designs in \cite{3,4,5} are not applicable to multicarriers systems with multiple FD transceivers.

Additionally,~deriving achievable rate regions for uplink and downlink via employing FD communication leads to an interesting trade-off problem which has attracted significant recent research works.
 As a result,~the rate region of multi-channel FD was investigated in \cite{8}~for two scenarios based on the fixed and general power allocation.~This problem was solved by a heuristic power allocation solution whereas analytical results were proposed for the fixed power allocation.
Furthermore,~the rate region of FD for the single and multi-channel was investigated in \cite{9}~where a heuristic algorithm was provided.~In contrast,~for HD communication systems,~both downlink and uplink are separated orthogonally in either time or frequency
and there is only trivial trade-off between the downlink and uplink.~Moreover, single-objective
optimization frameworks in \cite{3,4,5,6,Ata_G,Ata_T},~which were commonly adopted, are not suitable for FD systems,~due to the intrinsic coupling between uplink and downlink transmission
to study the throughput trade-off between uplink and downlink.~In addition, the works in \cite{8,9}~only proposed a heuristic solution while subchannel allocation was not addressed.

Motivated by the aforementioned discussions and in order to bridge the knowledge gap, in this paper,~we consider
two multi-objective optimization problems~(MOOPs)~which maximize the downlink and the uplink throughput in the first MOOP and minimize the total transmit power in the downlink
and uplink in the second MOOP,~respectively. We address these two MOOPs framework by the weighted Tchebycheff method.~Particularly,~each MOOP is transformed into a single objective optimization problem (SOOP) and then is solved suboptimally
by the majorization minimization~(MM)~approach~\cite{10,11,Ata_TWC}.
Moreover, we compare the performance of the IBFD communication system with that of the traditional HD and FD-HD systems where an FD-HD system consists of an FD BS serving HD users.
\begin{figure} \label{figSC}
  \centering
  \includegraphics[width=8.00cm,height=8.00cm]{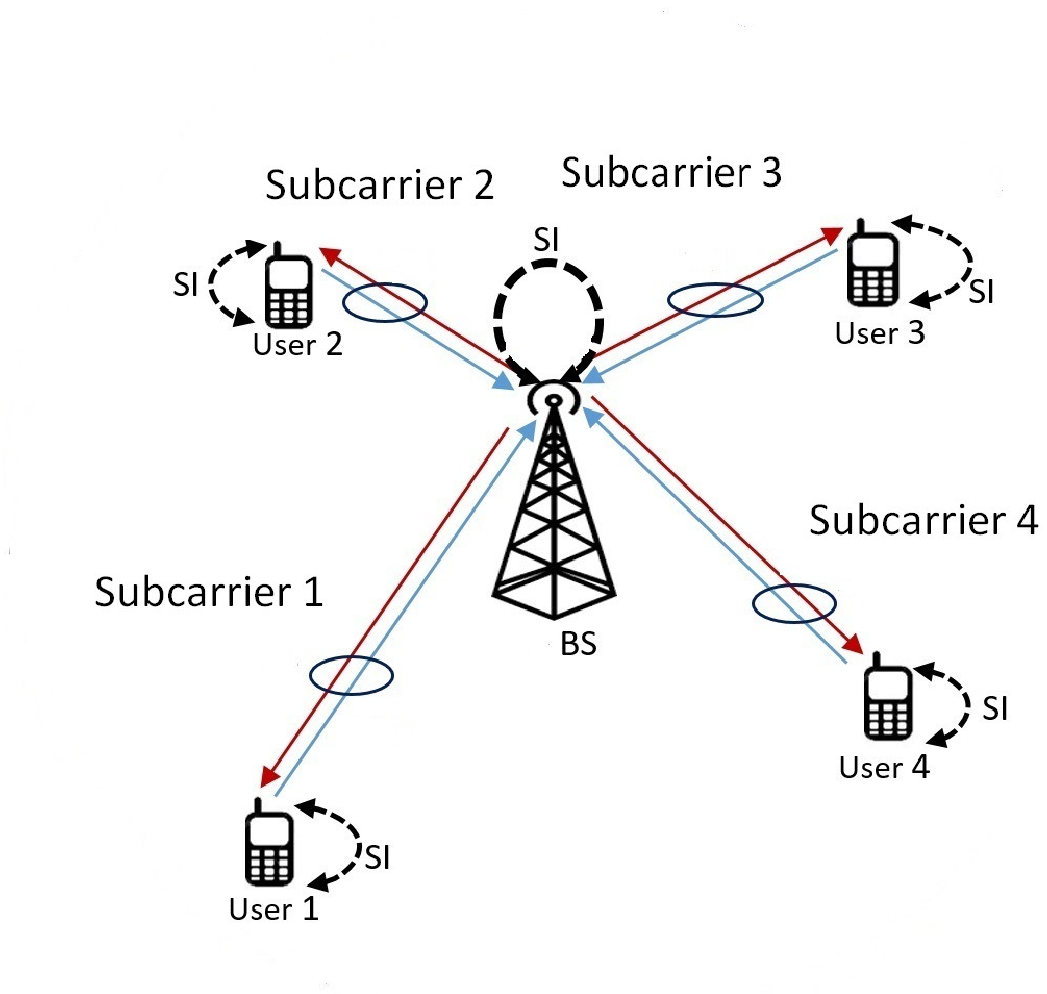}
 \caption{\small An IBFD-OFDMA system where the users and BS operate in FD mode.}
\end{figure}
 \section{System Model}
 We consider a single-cell network consisting of an FD BS and $K$ FD
users~(FD-FD) where the set of all users is defined as $\mathcal{K}=\{1,2,...,K\}$. The available bandwidth $W$ is divided into $M$ orthogonal frequency
division multiplexing (OFDM) subchannels for communications between the
users and the BS which is denoted by $\mathcal{M}=\{1,2,...,M\}$.
We further assume that all subcarriers are orthogonal to each
other without causing inter-subcarrier interference.~We consider that a subcarrier is allocated to a link for both uplink and
downlink directions.~Let us denote the subcarrier assignment as:
 \begin{equation*}
a_{m,k} = \begin{cases}
1, & \text{if subcarrier $m$ is allocated to user \textit{k}}, \\
0, & \text{otherwise}.
\end{cases}
\end{equation*}
Moreover, since the BS and the FD users are imperfect FD nodes,~and thus, suffer from non-negligible residual SI~\cite{Henggi,Azhang}.~In this system,~we consider different effective SI coefficients for the BS and users,~by constants~$\delta_{UE}$ and $\delta_{BS}$,~respectively~\cite{Henggi,Azhang,Mag,Ata_G} capturing the quality of self-interference cancellation (SIC) which is different in the users and the BS.
Also $q_{m,k}$ and $p_{m,k}$ denote the transmit power of user $k$ in subcarrier $m$ in the downlink and uplink, respectively.~Hence,~$q_{m,k}\times \delta_{BS}$, and $p_{m,k}\times \delta_{UE}$ represent the residual SI powers which are obtained by suppressing the simultaneous transmitting and receiving information signals after interference cancellation, respectively.~For notational simplicity,~we denote $\textbf{p}=[p_{m,k}]^{T}_{M\times K}$,~$\textbf{q}=[q_{m,k}]^{T}_{M\times K}$, and $\textbf{a}=[a_{m,k}]^{T}_{M\times K}$ as the vectors of power allocation for the users,~BS,~and subcarrier allocation,~respectively.~The sum rate in the downlink direction\footnote{Note that when FD is performed at the BS and the users operate in HD mode, the downlink data rate can be written as $R^{d}_{m,k}=\log_2(1+\frac{q_{m,k}h_{m,k}}{\sigma^2+p_{m,k}f_{m,k}})$, where $f_{m,k}$ denotes the co-channel interference between the UL user and the DL user.} is defined as
\begin{equation}
R^{d}(\textbf{a},\textbf{p},\textbf{q})=\sum_{k\in\mathcal{K}}\sum_{m\in\mathcal{M}}a_{m,k}R^{d}_{m,k},
\end{equation}
where $R^{d}_{m,k}=\log_2(1+\frac{q_{m,k}h_{m,k}}{\sigma^2+p_{m,k}\delta_{UE}})$
in which $h_{m,k}$ denotes the channel gain between the BS and user $k$ on sub-channel $m$.
Furthermore,~the sum rate in the uplink direction is given by
\begin{equation}
R^{u}(\textbf{a},\textbf{p},\textbf{q})=\sum_{k\in\mathcal{K}}\sum_{m\in\mathcal{M}}a_{m,k}R^{u}_{m,k},
\end{equation}
where $R^{u}_{m,k}=\log_2(1+\frac{p_{m,k}g_{m,k}}{\sigma^2+q_{m,k}\delta_{BS}})$ in which $g_{m,k}$ represents the channel gain between user $k$ and the BS on sub-channel $m$.
 The total transmission power for downlink and uplink are modeled as
		\begin{eqnarray}
		 P^{d}(\textbf{a},~\textbf{q})=
\sum_{k \in \mathcal{K}}\sum_{m\in \mathcal{M}} a_{m,k}\bigg(\frac{1}{\kappa}q_{m,k}\bigg)+P_c^\mathrm{BS},
\end{eqnarray}
and
\begin{eqnarray}
P^{u}(\textbf{a},~\textbf{p})=\sum_{k \in \mathcal{K}}\sum_{m\in \mathcal{M}}a_{m,k}\bigg(\frac{1}{\theta}p_{m,k}\bigg)
	+ K P_c^\mathrm{u},
	\end{eqnarray}
		 respectively,~where~$P_c^\mathrm{BS}$~and~$P_c^\mathrm{u}$  stand for circuit energy consumption of the BS and users, respectively. Furthermore, \\$0<\kappa,\theta<1$ are the power amplifier efficiency in the BS and the users,~respectively.
\section{Problem Formulation}
In this section, we investigate the performance trade-off between uplink and downlink in IBFD communication through formulating two MOOPs.~In general, it is desirable to design a subchannel assignment scheme as well as power allocation such that a given MOOP~(e.g.,~throughput trade-off or transmit power trade-off) is optimized subject to the feasibility of the transmit power level and the QoS requirement for both uplink and downlink directions.~To this end,~we formulate the design as the following optimization problem:
\begin{align}
\underset{\textbf{a},\textbf{p},\textbf{q}}{\mathrm{optimize}}& \quad U_{o}(\textbf{a},\textbf{p},\textbf{q})\\
\mathrm{s.t.} &~C_1:\underset{k\in\mathcal{K}}{\overset{}{\mathop \sum }}~ \underset{m\in\mathcal{M}}{\overset{}{\mathop \sum }}a_{m,k}~q_{m,k} \leq {P}^{\textrm{BS}}_{\max},\tag{5-1}\\
 &C_2:\underset{m\in\mathcal{M}}{\overset{}{\mathop \sum }}a_{m,k}~p_{m,k} \leq {P}^{\textrm{user}}_{\max},\tag{5-2}\\
&C_3:~p_{m,k}\geq 0,~C_4:~q_{m,k}\geq 0,\tag{5-3}\\
	& C_5:\underset{m\in\mathcal{M}}{\overset{}{\mathop \sum }} R_{m,k}^d \geq {R}_{\min}^d ,\tag{5-4}\\
	& C_6: \underset{m\in\mathcal{M}}{\overset{}{\mathop \sum }} R_{m,k}^u \geq {R}_{\min}^u,\tag{5-5}\\
&C_7:\underset{k\in\mathcal{K}}{\overset{}{\mathop \sum }} a_{m,k} \leq 1,\tag{5-6}\\
&C_8:~a_{m,k} \in \{0,1\},\tag{5-7}
\end{align}
where~$U_{o}(\textbf{a},\textbf{p},\textbf{q})$~is a certain objective function which will be defined for each problem later.

In problem (5),~constraints $C_1$ and $C_2$ put a limit on the maximum transmit power for BS~(${P}^{\textrm{BS}}_{\max}$) and user~(${P}^{\textrm{user}}_{\max}$),~respectively.~Furthermore, $C_3$ and $C_4$ guarantee the non-negativity of the power allocation variables.~Constraints $C_5$ and $C_6$ guarantee the data rate QoS in downlink~(${R}_{\min}^d$)~and uplink~(${R}_{\min}^u$),~respectively.~Constraint $C_7$~states that each subchannel can only be allocated to at most one user~and constraint $C_8$ indicates that the sub-channel allocation variable is binary.

$U_{o}$~in the first MOOP is formulated to maximize jointly the total downlink and uplink data rate as follows.\\

\textit{Problem 1} (MOOP Framework for Throughput):
\begin{eqnarray}\label{5}
\begin{aligned}
&\max_{\textbf{a},\textbf{p},\textbf{q}}~~ R^{d}(\textbf{a},\textbf{p},\textbf{q}) \\
&\max_{\textbf{a},\textbf{p},\textbf{q}}~~ R^{u}(\textbf{a},\textbf{p},\textbf{q}) \\
\text{s.t.} ~&C_{1}-C_{8}.
\end{aligned}
\end{eqnarray}
On the other hand,~the second set of desirable objective functions aim to minimize the total downlink transmission power and uplink
transmission power, simultaneously.~$U_{o}$ in the second MOOP is given by
\\

\textit{Problem 2} (MOOP Framework for Transmit Power):
\begin{eqnarray}\label{joint_power}
\begin{aligned}
 &\min_{\textbf{a},\textbf{q}}{P}^{d}(\textbf{a},\textbf{q})\\
 &\min_{\textbf{a},\textbf{p}}{P}^{u}(\textbf{a},\textbf{p})\\
\text{s.t.} ~&C_{1}-C_{8}.
\end{aligned}
\end{eqnarray}
The two aforementioned problems are desirable and their solutions are the key to unleash the performance of the considered system.~Yet,~in IBFD wireless communication systems,~these objectives in each MOOP conflict with each other, due to the SI caused by the transmit power in FD operation.
To tackle this issue,~we exploit the~MOOP which is commonly employed to study the trade-off between competing objective functions~\cite{12,Ata_G,Ata_WCNC}.
To obtain the Pareto optimal solution,~a new optimization problem is formulated to investigate the trade-off between downlink and uplink by employing the weighted Tchebycheff method~\cite{7,Ata,Ata_TWC}.~It is worthwhile to note that the weighted Tchebycheff method can achieve every Pareto optimal solution with much a lower computational complexity~\cite{12,Ata,Ata_TWC} compared to other existing approaches.~The complete Pareto optimal set can be achieved by solving the following multi-objective problem:\\


\textit{Transformed Problem 1}~(MOOP to SOOP for Throughput Trade-off):
\begin{align}
&\min_{\textbf{a},\textbf{p},\textbf{q},\psi}~\psi\nonumber\\
\text{s.t.}~
&\nu_{n}(U^{*}_{n}-U_n(\textbf{a},\textbf{p},\textbf{q}))\leq \psi,\nonumber\\
 &C_{1}-C_{8},~\forall{n\in\{1,2\}},
\end{align}
 where $U_1(\textbf{a},\textbf{p},\textbf{q})=R^{d}(\textbf{a},\textbf{p},\textbf{q})$ and $U_2(\textbf{a},\textbf{p},\textbf{q})=R^{u}(\textbf{a},\textbf{p},\textbf{q})$.~The second transformed problem is formulated as follow:\\

\textit{Transformed Problem 2}~(MOOP to SOOP for Transmit Power Trade-off):
\begin{align}
&\min_{\textbf{a},\textbf{p},\textbf{q},\chi}~\chi\nonumber\\
\text{s.t.}~
&\nu_{n}(U_n(\textbf{a},\textbf{p},\textbf{q})-U^{*}_{n})\leq \chi,\nonumber\\
 &C_{1}-C_{8},~\forall{n\in\{3,4\}},
\end{align}
where $U_3(\textbf{a},\textbf{q})=p^{d}(\textbf{a},\textbf{q})$ and $U_4(\textbf{a},\textbf{p})=p^{u}(\textbf{a},\textbf{p})$.~Furthermore,~$\nu_{n}\geq 0$, $\sum_{n}\nu_{n}=1$~denotes the weighting coefficient reflecting the importance of different objectives.~Moreover,~$\psi$ and $\chi$ are auxiliary optimization variables.~Since \textit{Problem 1} and \textit{Problem 2} share the similar structure,~we only focus on the suboptimal solution development of transformed \textit{Problem 1} in the sequel\footnote{The solution of \textit{Problem 2} can be obtained by applying the resource allocation scheme designed for \textit{Problem 1}.}.
\section{Proposed solution}
In order to obtain the Pareto optimal boundary,~we propose a method to find a solution for each problem ($U^{*}_{n}$).
One can readily verify that problems (8) and (9) are non-convex mixed-integer nonlinear programming~(MINLP)~due
 to the interference coupled in the rate functions and the discrete nature of subchannel allocation.
 In order to facilitate the design of computationally-efficient solution,~we first deal with non-convex constraints $C_{1}$~and~$C_{2}$ by first introducing
following constraints
\begin{align}
       &C^{'}_{1}:~q_{m,k}\le a_{m,k} \ {P}^{\textrm{BS}}_{\max},\underset{k\in\mathcal{K}}{\overset{}{\mathop \sum }}~ \underset{m\in\mathcal{M}}{\overset{}{\mathop \sum }}q_{m,k} \leq {P}^{\textrm{BS}}_{\max},~\\
                            &\textrm C^{'}_{2}:~p_{m,k}\le a_{m,k} \ {P}^{\textrm{user}}_{\max},~\underset{m\in\mathcal{M}}{\overset{}{\mathop \sum }}p_{m,k} \leq {P}^{\textrm{user}}_{\max}.
  \end{align}
Furthermore,~we adopt a similar approach as in \cite{6,11,Ata_WCL}~to handle the binary variable constraint in~$C_{8}$ and rewrite this constraint into the following equivalent form:
\begin{eqnarray}\label{18,19}
\begin{aligned}
  &\mathcal{R}_1:0\leq a_{m,k}\leq 1,~\\
  &\mathcal{R}_2:\sum_{k}\sum_{m}a_{m,k}-\ (a_{m,k})^2 \leq 0.
\end{aligned}
\end{eqnarray}
   Then,~we introduce a penalty factor $\lambda$~and optimization problem (8) is equivalent to the following problem
    for a large value~$\lambda\gg1$
    \cite{6},~\cite{Ata_TWC}:
  \begin{eqnarray}\label{12}
 \begin{aligned}
 & \min_{\textbf{a},\textbf{p},\textbf{q},\psi}\psi
  +\lambda\Big(E(\textbf{a})-D(\textbf{a})\Big)\\ &\text{s.t.}~~
  \nu_{n}\big (U^{*}_{n}-U_n(\textbf{a},\textbf{p},\textbf{q})\big)\leq \psi,\\
  &C^{'}_1,C^{'}_2,C_{3}-C_{4},C_{7},\mathcal{R}_{1},\\
  &C_5: F^{d}(\textbf{p},\textbf{q})-G^{d}(\textbf{p})\geq {R}_{\min}^d ,~\forall{k}\in \mathcal{K},\\
  &C_6: F^{u}(\textbf{p},\textbf{q})-G^{u}(\textbf{q})\geq {R}_{\min}^u ,~\forall{k}\in \mathcal{K},
    \end{aligned}
  \end{eqnarray}
  where, $F^{d}(\textbf{p},\textbf{q})$,~$G^{d}(\textbf{p})$,~$F^{u}(\textbf{p},\textbf{q})$,~$G^{u}(\textbf{q})$,~$E(\textbf{a})$,~and~$D(\textbf{a})$~are defined as follows
\begin{align}
  &F^{d}(\textbf{p},\textbf{q})\triangleq\underset{m\in\mathcal{M}}{\overset{}{\mathop \sum }} \log_2\big({q_{m,k} h_{m,k}}+\sigma^2+p_{m,k} \delta_{UE} \big),\\
  &G^{d}(\textbf{p})\triangleq\underset{m\in\mathcal{M}}{\overset{}{\mathop \sum }} \log_2(\sigma^2+p_{m,k} \delta_{UE} \big),\\
  &F^{u}(\textbf{p},\textbf{q})\triangleq \underset{m\in\mathcal{M}}{\overset{}{\mathop \sum }}\log_2\big({p_{m,k} g_{m,k}}+\sigma^2+q_{m,k} \delta_{BS} \big),\\
  &G^{u}(\textbf{q})\triangleq \underset{m\in\mathcal{M}}{\overset{}{\mathop \sum }}\log_2\big(\sigma^2+q_{m,k} \delta_{BS} \big),\\
  &\small E(\textbf{a})\triangleq\underset{k}{\overset{}{\mathop \sum }}\underset{m}{\overset{}{\mathop \sum }}a_{m,k},~
  \small D(\textbf{a})\triangleq\underset{k}{\overset{}{\mathop \sum }}\underset{m}{\overset{}{\mathop \sum }}\Big(a_{m,k}\Big)^2,
\end{align}
   respectively.~In fact, $\lambda$ serves as a penalty
factor to penalize the objective function when $a_{m,k}$
 is not binary values.~The resulting optimization problem (\ref{12})~is still non-convex.
  To facilitate the solution design,~we apply~the MM~approach to approximate (13) by a convex optimization problem using the first order Taylor approximation~\cite{10}.~Since $G^{d}(\textbf{p})$,~$G^{u}(\textbf{q})$,~and $D(\textbf{a})$ are differentiable convex functions,~for any feasible points $\textbf{p}^{t}$,~$\textbf{q}^{t}$,~and $\textbf{a}^{t}$,~the following inequalities hold:
 \begin{align}
    &G^{d}(\textbf{p})\leq G^{d}(\textbf{p}^{t})
 +\nabla_{{\textbf{p}}}G^{d}({\textbf{p}}^{t})({\textbf{p}}-{\textbf{p}}^{t})\triangleq \tilde{G}^{d}({\textbf{p}}),\\
     &G^{u}(\textbf{q})\leq G^{u}(\textbf{q}^{t})
 +\nabla_{{\textbf{q}}}G^{u}({\textbf{q}}^{t})({\textbf{q}}-{\textbf{q}}^{t})\triangleq \tilde{G}^{u}({\textbf{q}}),\\
 &{D}(\textbf{a})\geq D(\textbf{a}^{t})
 +\nabla_{\textbf{a}}D(\textbf{a}^{t})(\textbf{a}-\textbf{a}^{t})\triangleq \tilde{D}(\textbf{a}),
 \end{align}
 where ~$\nabla_{\textbf{p}}$,~$\nabla_{\textbf{q}}$,~and $\nabla_{\textbf{a}}$ denote the gradient operation with respect to $\textbf{p}$,~$\textbf{q}$,~and $\textbf{a}$,~respectively.~It can be seen that $\tilde{G}^{d}({\textbf{p}})$,~$\tilde{G}^{u}({\textbf{q}})$,~and $\tilde{D}(\textbf{a})$ are all affine functions.~Therefore for any given $\textbf{p}^{t}$,~$\textbf{q}^{t}$,~and $\textbf{a}^{t}$, we solve the following convex optimization problem:

\begin{eqnarray}\label{Main_Trade}
\begin{aligned}
&\min_{\textbf{a},\textbf{p},\textbf{q},\psi}~\psi +\lambda\Big(E(\textbf{a})-\tilde{D}(\textbf{a})\Big) \\
\text{s.t.}~&\nu_{n}\big (U^{*}_{n}-\hat {U}_n(\textbf{p},\textbf{q})\big)\leq \psi,\\
	&C^{'}_1,C^{'}_2,C_{3}-C_{4},C_{7},\mathcal{R}_{1},\\
&C_5:F^{d}(\textbf{p},\textbf{q})-\tilde{G}^{d}(\textbf{p})\geq {R}_{\min}^d ,~\forall{k}\in \mathcal{K},\\
&C_6:F^{u}(\textbf{p},\textbf{q})-\tilde{G}^{u}(\textbf{q})\geq {R}_{\min}^u ,~\forall{k}\in \mathcal{K},
\end{aligned}
\end{eqnarray}
where~$U^{*}_{n}$~can be obtained from each objective,~respectively.~Note that $\hat {U}_n(\textbf{p},\textbf{q})$ is the convexified objective function\footnote{Note that $\hat {U}_1=\hat{R}^{d}(\textbf{p},\textbf{q})=\sum_{k\in \mathcal{K}}F^{d}(\textbf{p},\textbf{q})-\tilde{G}^{d}(\textbf{p})$ and $\hat {U}_2=\hat{R}^{u}(\textbf{p},\textbf{q})=\sum_{k\in \mathcal{K}}F^{u}(\textbf{p},\textbf{q})-\tilde{G}^{u}(\textbf{q}).$}~for data rate in downlink and uplink.~Since the optimization problem in (\ref{Main_Trade}) is convex~and satisfies the Slater's constraint qualification \cite{14}, it can be effectively solved using optimization packages incorporating interior-point methods such as CVX~\cite{13} with polynomial time computational complexity.~Note that solving (\ref{Main_Trade}) for a given set of $\textbf{p}^{t}$,~$\textbf{q}^{t}$,~and $\textbf{a}^{t}$ results in an upperbound value of (13).  Therefore,~an iterative algorithm can be employed to tighten the obtained upper bound where the solution of (\ref{Main_Trade}) in iteration $(t)$ is used as an initial point for the next iteration ($t+1$) \cite{Ata_TWC}.~It is noteworthy that the sub-optimal iterative algorithm obtains a locally optimal solution of the original problem (13) with a polynomial time complexity~\cite{6,11,Ata_TWC}.~The solution methodology for handling optimization problem (\ref{Main_Trade}) is summarized in \textbf{Algorithm} 1.

  \begin{algorithm}[t]\label{algorithm3}
 \caption{\small Proposed MOOP algorithm based on majorization-minimization~(MOOP-MM)}
 \begin{algorithmic}[1]
 \State~Set $t=0$ and initialize the maximum number of iteration $T_{\max}$,
 \State Set penalty factor $\lambda\gg1$ and proper weighting coefficient factor $(\nu)$
 \State Set feasible set vector $\textbf{p}^{0} $,~$\textbf{q}^{0}$, and $\textbf{a}^{0}$
\State{\textbf{~Repeat}}
\State~Update $\tilde{G}^{d}({\textbf{p}})$,~$\tilde{G}^{u}({\textbf{q}})$,~and $\tilde{D}(\textbf{a})$ as presented in~({19}),~({20}),~and ({21}),~respectively.
\State Obtain $U^{*}_n$ through solving each objective,~independently.
\State Solve optimization problem of (\ref{Main_Trade}) and save the intermediate resource allocation policy~$\textbf{p}^{t}$,~${\textbf{q}^{t}}$, and $\textbf{a}^{t}$
\State~Set $t=t+1$ and $\textbf{p}^{t}=\textbf{p}$,~$\textbf{q}^{t}=\textbf{q}$,~and $\textbf{a}^{t}=\textbf{a}$
\State \textbf{~Until }convergence or $t=T_{\max}$
\State~$\textbf{p}^{*}=\textbf{p}^{t}$,~$\textbf{q}^{*}=\textbf{q}^{t}$,~$\textbf{a}^{*}=\textbf{a}^{t}$
\end{algorithmic}
\end{algorithm}
Similar to the method for studying the throughput trade-off, we also apply the weighted
Tchebycheff method to transform the MOOP into a SOOP to investigate the trade-off between downlink and uplink transmission power which is given as below
\begin{eqnarray}
\begin{aligned}
&\min_{\textbf{a},\textbf{p},\textbf{q},\chi}~\chi +\lambda\Big(E(\textbf{a})-\tilde{D}(\textbf{a})\Big) \\
 \text{s.t.}~&\nu\big (p^{d}(\textbf{a},\textbf{q})-p^{d}_{\min}\big)\leq \chi,\\
&(1-\nu)\big (p^{u}(\textbf{a},\textbf{p})-p^{u}_{\min}\big)\leq \chi,\\
	&C^{'}_1,C^{'}_2,C_{3}-C_{4},C_{7},\mathcal{R}_{1},\\
&C_5:F^{d}(\textbf{p},\textbf{q})-\tilde{G}^{d}(\textbf{p})\geq {R}_{\min}^d ,~\forall{k}\in \mathcal{K},\\
&C_6:F^{u}(\textbf{p},\textbf{q})-\tilde{G}^{u}(\textbf{q})\geq {R}_{\min}^u ,~\forall{k}\in \mathcal{K},
 \end{aligned}
\end{eqnarray}
where  $p^{d}_{\min}$ and $p^{u}_{\min}$ are the optimal points as the total downlink and uplink minimization of each objective,~respectively.
\section{Simulation results}
In this section, the performance of the
proposed MOOP based on the resource allocation
approach is studied via extensive simulations.~For the wireless channel
model, each subchannel experiences both Rayleigh fading and pathloss as in~\cite{6}.~There are 10 users randomly distributed in the network.~Without loss of generality, we assume that the maximum transmit power budget for each user is 23~dBm and for the BS is 42~dBm.~Other important simulation parameters are given in Table I, unless otherwise is specified. 
\begin{table}
\caption{Simulation Parameters}
\label{Simulation Parameters}
\centering
\begin{tabular}{|c|c|}\hline
{\bf Parameter} & {\bf Value} \\ \hline
{Cell radius} & $100$ m \\ \hline
Number of users in each cell  & $10$\\ \hline
Number of sub-channels, $M$& $32$ \\ \hline
Maximum transmit power of user &$p^{u}_{\max}=23$ dBm\\ \hline
Maximum transmit power of BS &$p^{d}_{\max}=42$ dBm\\ \hline
$\delta_{BS}=\delta_{UE}$& $-90$ dB\\ \hline
Power amplifier efficiency of the BS ($\kappa$)&$38/100$\\ \hline
Power amplifier efficiency of the UE ($\psi$)&$20/100$\\ \hline
$R^{d}_{\min}$ & $10$ bps/Hz\\ \hline
$R^{u}_{\min}$ & $5$ bps/Hz\\ \hline
$\sigma^{2}$& $-120~\textrm{dBm}$\\ \hline
 Bandwidth of each subcarrier &$180$ kHz\\ \hline
Total number of channel realization & $50$\\ \hline
$\lambda$& $10^{\log(\frac{{P}^{\textrm{BS}}_{\max}}{\sigma^{2}})}$\\
\hline
\end{tabular}
\end{table}
\subsection{Data Rate Trade-off Region}
Fig.~2 illustrates the trade-off between the downlink and uplink total data rate for different values of residual SI.
This figure is achieved by solving problem (22) for different values of $\nu_{n} \in [0,1]$,~$n\in \{1,2\}$, with a step size of $0.05$ such that~$\sum_{n}\nu_{n}=1$.
It can be observed that there is a non-trivial trade-off between the total data rate of downlink and uplink.
In particular,~the total uplink data rate is a monotonically decreasing function versus the total downlink data rate.~In other words, maximizing the total uplink data rate results in a reduction on total downlink data rate as these objectives are conflicting with each other.
Moreover,~this figure demonstrates that when the SI becomes more severe,~system throughput in both uplink and downlink decreases.~This observation can be explained by the fact that as the residual SI increases, the interference term in the rate function increases as well which results in a degradation in the system throughput.~This figure also shows that the data rate for uplink and downlink decreases as the minimum data rate requirement keeps increasing.~In fact when the minimum required data rate increases,~the resource allocator has become less flexible in allocating subcarriers.~For example,~some
of the subcarriers that were previously assigned to users
with high channel quality, would be allocated to users with
deep faded channel gains to satisfy the more stringent data rate requirement.~In other words, the system resources are underutilized leading to a reduction of data rate.

For comparison, we also consider two other baseline schemes.~For baseline scheme 1,~we adopt the simulation based on the HD mode where the frequency band adopted for uplink communications is orthogonal to that of the downlink direction.
This figure shows that although the achievable data rate in HD mode is not impaired by the SI,
the total system throughput is significantly less than that of IBFD.
This is due to the fact that FD communication in IBFD systems has the potential to double the spectral efficiency via a more flexible spectrum utilization which leads to a significantly higher data rate as compared to HD mode.
For baseline scheme 2, we focus on the FD-HD protocol where the system includes an FD BS serving $K$ HD users.
As observed,~the proposed scheme achieves a higher system throughput compared to other considered schemes due to optimized resource allocation for simultaneous transmission uplink and downlink directions over the same frequency band.
In addition,~the inherent co-channel interference in FD-HD\footnote{Note that for FD-HD, the optimization is formulated as the one in \cite{6} adopting the OFDMA scheme where all the users are equally divided for uplink and downlink.}
systems degrades both uplink and downlink rate.
Hence, one can conclude that even for moderate level of SI,~enabling FD communication of both BS and users, can improve the system performance substantially.
\begin{figure} \label{figSC}
  \centering
  \includegraphics[width=9.500cm,height=7.00cm]{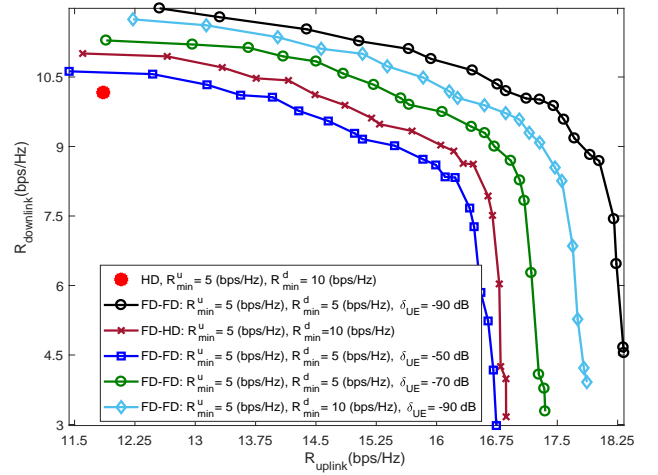}
  \caption{\small Throughput trade-off region of downlink and uplink.}
\end{figure}
\subsection{Power Transmission Trade-off Region}
In Fig.~3 the power consumption trade-off region between the downlink and uplink transmission is investigated.
Similar to the throughput trade-off,~the uplink and downlink transmit power minimization conflict with each other.~Specifically,~the imperfect of downlink and transmit powers are coupled via the SI which leads to a non-trivial trade-off between uplink and downlink communications.~In fact, increasing the downlink transmit power also leads to a high power level of residual SI jeopardizing the uplink data rate.
Furthermore,~the FD-HD trade-off region is substantially smaller than that of the FD-FD system.~In other words,~the proposed FD-FD is generally more power efficient as it can provide a higher flexibility in resource allocation.~In fact,~the performance of FD-FD system is always superior to the FD-HD system for the considered practical values of $\delta_{UE}$.
\begin{figure} \label{figSC}
  \centering
  \includegraphics[width=9.500cm,height=7.00cm]{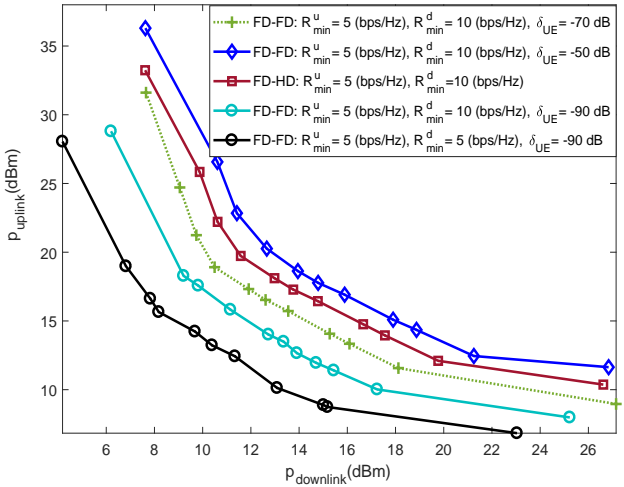}
  \caption{\small Power consumption trade-off region of uplink and downlink for different values of weights.}
\end{figure}
\section{conclusion}
In this paper,~we investigated the performance trade-off between downlink and uplink in a single-cell IBFD-OFDMA-based networks.
The algorithm design was formulated by two MOOPs that jointly optimizes the uplink and downlink for data rate in the first MOOP and aggregate power consumption in the second MOOP,~respectively.~The MOOP was handled by the weighted Tchebycheff method and solved suboptimally via the MM approach by constructing a sequence of surrogate function.
 Simulation results demonstrated the trade-off between the
studied competing objective functions.
In particular, our results indicated the superiority of the proposed method to conventional HD-OFDMA as well as FD-HD OFDMA for practical values of residual SI.


\begin{thebibliography}{1}
\bibitem{1}V. W. Wong, R. Schober, D. W. K. Ng, and L.C.Wang, \textit{Key Technologies
for 5G Wireless Systems.} Cambridge University Press, 2017.

\bibitem{Multiple_Antenna_5G} J. Zhang, E. Björnson, M. Matthaiou, D. W. K. Ng, H. Yang, D. J. Love ``Multiple Antenna Technologies for Beyond 5G" Sep.~2019.~[Online].~Available:~http://arXiv:1910.00092v1
\bibitem{Massive_Access_5G} X. Chen, D. W. K. Ng, W. Yu, E. G. Larsson, N. Al-Dhahir, R. Schober ``Massive Access for 5G and Beyond"
Feb.~2020.~[Online].~Available:~http://arXiv:2002.03491v1
\bibitem{2} A. Sabharwal, P. Schniter, D. Guo, D. W. Bliss, S. Rangarajan, and
R. Wichman, ``In-band full-duplex wireless: Challenges and opportunities,"
\textit{IEEE J. Select. Areas Commun.}, vol. 32, no. 9,
pp. 1637-1652, Sep.~2014.
\bibitem{3}J. H. Yun, ``Intra and inter-cell resource management in full-duplex
heterogeneous cellular networks," \textit{IEEE Trans. Mobile Comput.,}~vol. 15, no. 2, pp. 392-405, Feb. 2016.

\bibitem{4}C. Nam, C. Joo, and S. Bahk, ``Joint subcarrier assignment
and power allocation in full-duplex OFDMA networks," \textit{IEEE
Trans. Wireless Commun,} vol. 14, no. 6, pp.
3108-3119, Jun.~2015.
\bibitem{5}S. Zarandi, M. Rasti, ``Resource allocation in inband full-duplex two-tier networks with quality of service provisioning", \textit{Proc.~IEEE WCNC}, pp. 1-6, Jun. 2018

    \bibitem{6}Y. Sun, D. W. K. Ng, Z. Ding, and R. Schober, ``Optimal joint power and subcarrier allocation for full-duplex multicarrier
non-orthogonal multiple access systems," \textit{IEEE Trans. Commun.,} vol. 65, no. 3, pp. 1077-1091, Mar. 2017.

\bibitem{Ata_G} A. Khalili, S. Zarandi, M. Rasti,~and E. Hossain ``Multi-objective optimization for energy-and spectral-efficiency tradeoff in in-band full-duplex (IBFD) communication" Accepted by \textit{IEEE Global Communications Conference} Jul.~2019.~[Online].~Available:~http://arXiv:1907.08250v1

\bibitem{Ata_T}R. Aslani, M. Rasti, and A. Khalili, ``Energy efficiency maximization via joint sub-carrier assignment and power control for OFDMA full duplex networks," \textit{IEEE Trans.~Veh.~Technol.,}~vol. 68, no. 12, pp. 11859-11872, Dec. 2019.
\bibitem{7}Y. Sun, D. W. K. Ng, J. Zhu, and R. Schober, ``MultiObjective optimization for robust power efficient and secure full-duplex
wireless communication systems," \textit{IEEE Trans. Wireless Commun.}, vol. 15, no. 8, pp. 5511-5526, Apr. 2016.

\bibitem{8}W. Li, J. Lilleberg, and K. Rikkinen, ``On rate region analysis of half- and
full-duplex OFDM communication links," \textit{IEEE J. Sel. Areas Commun.,}
vol. 32, no. 9, pp. 1688-1698, Sep. 2014.
\bibitem{9}J. Diakonikolas and G. Zussman, ``On the rate regions of single-channel and multi-channel full-duplex links,"
 \textit{IEEE/ACM Trans. Netw.,} vol. 26, no. 1, pp. 47-60, Feb. 2018.
\bibitem{Azhang}D. Ramirez and B. Aazhang, ``Optimal routing and power allocation for wireless networks with imperfect full-duplex nodes," \textit{IEEE Trans.~Wireless Commun.,} vol. 12, no. 9, pp. 4692-4704, Sep.~2013.

\bibitem{Henggi}Z. Tong and M. Haenggi, ``Throughput analysis for full-duplex wireless networks with imperfect self-interference cancellation," \textit{IEEE Trans. Commun.,} vol. 63, no. 11, pp. 4490-4500, Nov. 2015.

\bibitem{Mag}R. Li, Y. Chen, G. Y. Li, and G. Liu, ``Full-duplex cellular networks," \textit{IEEE Commun. Mag.,} vol. 55, no. 4, pp. 184-191, Apr.~2017.

\bibitem{10}Y. Sun, P. Babu, and D. P. Palomar, ``Majorization-minimization algorithms in signal processing, communications, and
machine learning," \textit{IEEE Trans. Signal Process.,} vol. 65, no. 3, pp. 794-816, Feb. 2017.

\bibitem{11} A. Khalili, S. Zarandi, and M. Rasti, ``Joint resource allocation and offloading decision in mobile edge computing," \textit{IEEE Commun. Letts.,}~vol. 23, no. 4, pp. 684-687, Apr. 2019.

\bibitem{12}K. Miettinen, \textit{Nonlinear Multiobjective Optimization.} Springer, 1999.

\bibitem{Ata}M. R. Mili, A. Khalili, D. W. K. Ng, and H. Steendam, ``A novel performance tradeoff in heterogeneous networks: A multi-objective approach," \textit{IEEE Wireless Commun. Letts.,} vol. 8, no. 5, pp. 1402-1405, Oct. 2019.

\bibitem{Ata_TWC} A. Khalili, M. R. Mili, M. Rasti, S. Parsaeefard, and D. W. K. Ng, ``Antenna selection strategy for energy efficiency maximization in uplink OFDMA networks: A multi-objective approach," \textit{IEEE Trans.~Wireless Commun.}~vol. 19, no. 1, pp. 595-609, Jan. 2020.

\bibitem{Ata_WCL}A. Khalili, S. Akhlaghi, H. Tabassum and D. W. K. Ng, ``Joint user association and resource allocation in the uplink of heterogeneous networks," \textit{IEEE Wireless Commun. Lett.} Early Access. Jan.~2020.

\bibitem{Ata_WCNC} A. Khalili and D. W. K. Ng ``Energy and spectral efficiency tradeoff in OFDMA networks via antenna selection strategy" Accepted by \textit{IEEE Wireless Communications and Networking Conference (WCNC)} Jan.~2020.~[Online].~Available:~http://arXiv:2002.04104v1

\bibitem{14}S. Boyd and L. Vandenberghe, \textit{Convex Optimization.} Cambridge
University Press, 2004.

\bibitem{13}M. Grant and S. Boyd, ``CVX: Matlab software for disciplined convex
programming, version 2.1," http://cvxr.com/cvx, Mar. 2014.

\end{thebibliography}
\end{document}